# Fermiology of $Sr_4V_2O_6Fe_2As_2$: Quasi-Nested Fe *vs* Mott-Insulating V Orbitals

T. Qian[1], N. Xu[1], Y.-B. Shi[1], K. Nakayama[2], P. Richard[1,3], T. Kawahara[2], T. Sato[2,4], T. Takahashi[2,3], M. Neupane[5], Y.-M. Xu[5], X.-P. Wang[1], G. Xu[1], X. Dai[1], Z. Fang[1], P. Cheng[1], H.-H. Wen[1], and H. Ding[1]

[1] *Beijing National Laboratory for Condensed Matter Physics, and Institute of Physics, Chinese Academy of Sciences, Beijing 100190, China*

[2] *Department of Physics, Tohoku University, Sendai 980-8578, Japan*

[3] *WPI Research Center, Advanced Institute for Materials Research, Tohoku University, Sendai 980-8577, Japan*

[4] *TRiP, Japan Science and Technology Agency (JST), Kawaguchi 332-0012, Japan*

[5] *Department of Physics, Boston College, Chestnut Hill, MA 02467, USA*

We have performed an angle-resolved photoemission spectroscopy study of a new iron-based superconductor $Sr_4V_2O_6Fe_2As_2$. While V 3*d* orbitals are found to be in a Mott insulator state and show an incoherent peak at ~ 1 eV below the Fermi level, the dispersive Fe 3*d* bands form several hole- and electron-like Fermi surfaces (FSs), some of which are quasi-nested by the ($\pi$, 0) wave vector. This differs from the local density approximation (LDA) calculations, which predict non-nested FSs for this material. However, LDA+U with a large effective Hubbard energy U on V 3*d* electrons can reproduce the experimental observation reasonably well. The observed fermiology in superconducting $Sr_4V_2O_6Fe_2As_2$ strongly supports that ($\pi$, 0) interband scattering between quasi-nested FSs is indispensable to superconductivity in pnictides.

The most important question for superconducting (SC) pnictides is how the low-energy electrons are paired in these materials. Since the SC pairing occurs in the vicinity of the Fermi level ($E_F$), the low-energy band structure and the fermiology are critical to the pairing. Angle-resolved photoemission spectroscopy (ARPES) showed that all SC pnictides share similar fermiology, where hole-like Fermi surfaces (FSs) near the Brillouin zone (BZ) center ($\Gamma$) are quasi-nested to electron-like FSs near the M point by the ($\pi$, 0) wave vector [($\pi$, $\pi$) in the reconstructed BZ] [1-3]. Inelastic neutron scattering measurements have revealed the presence of spin fluctuations near ($\pi$, 0) in several families of pnictides [4-6]. These experimental results confirm the importance of the interband scattering by the ($\pi$, 0) spin fluctuations in the pairing, suggesting the universality of pairing interactions in the pnictides.

However, the universality has encountered a serious challenge in the new iron-based superconductor $Sr_4V_2O_6Fe_2As_2$ (called "42622") [7-9], since most local density approximation (LDA) band calculations [10-13], which have been quite reliable in calculating the FS of other pnictides, predicted that this new pnictide has no good quasi-nested FS segments. $Sr_4V_2O_6Fe_2As_2$ can be regarded as a superlattice consisting of alternating stacking of $SrFe_2As_2$ (called "122") and perovskite-like $Sr_3V_2O_6$ layers. Compared with other pnictides, the most distinctive characteristics of $Sr_4V_2O_6Fe_2As_2$ predicted by LDA calculations is the presence of metallic V 3$d$ bands. Although the V 3$d$ bands weakly couple to the Fe 3$d$ bands, small hybridization near $E_F$ between them changes the FS topology and destroys the quasi-nesting condition [10]. It is nevertheless argued by one of the LDA papers [11] that if only the Fe-derived FS system is taken into account, the bare susceptibility shows a peak at ($\pi$, 0) similar to other pnictides. Furthermore, it has been pointed out that the V 3$d$ electrons in $Sr_4V_2O_6Fe_2As_2$ may be subject to strong on-site electron correlations that would remove V 3$d$ states from $E_F$ [11]. It is thus of particular importance to investigate the electronic structure and FS by performing ARPES measurements on $Sr_4V_2O_6Fe_2As_2$.

In this Letter, we present ARPES results on $Sr_4V_2O_6Fe_2As_2$. We find that the V 3$d$ orbitals are in a Mott insulating state and show an incoherent peak, or the lower Hubbard band (LHB), at a binding energy ($E_B$) of ~ 1 eV. All the dispersive bands within 0.8 eV below $E_F$ can be mostly attributed to Fe 3$d$ states. The observed FS topology is essentially similar to those of other pnictide superconductors. The FSs are quasi-nested by the (π, 0) wave vector, suggesting the importance of the (π, 0) interband scattering for the pairing in $Sr_4V_2O_6Fe_2As_2$.

The high-quality single crystals of $Sr_4V_2O_6Fe_2As_2$ were grown by the flux method. ARPES measurements were performed at beamline 28A of the Photon Factory (KEK, Tsukuba), beamline U-71A of the Synchrotron Radiation Center (Stoughton, WI), and Tohoku University using the He I$\alpha$ resonance line ($h\nu$ = 21.2 eV). The energy and angular resolutions were set to 30 meV and 0.2°, respectively. Samples with the tiny size of < 0.2 × 0.2 mm$^2$ were carefully mounted using a robotic mounting device, cleaved *in situ* and measured at 40 K in a working vacuum better than 1 × 10$^{-10}$ Torr. The $E_F$ of the samples was referenced to that of a gold film evaporated onto the sample holders.

As pointed out by LDA calculations, the valence of V ions is important to the fermiology of $Sr_4V_2O_6Fe_2As_2$. To determine the valence of V in $Sr_4V_2O_6Fe_2As_2$, we have measured a core-level spectrum at $h\nu$ = 220 eV, as shown in Fig. 1(a). We identify a broad V 3$s$ peak at $E_B$ ~ 70 eV and a V 3$p$ peak whose leading edge at the lower-$E_B$ side is at 37.2 eV. The broad lineshape of the 3$s$ peak in $Sr_4V_2O_6Fe_2As_2$ is more similar to those in $V_2O_3$ and $VO_2$ than the one in $V_2O_5$, which is much narrower due to the disappearance of the multiple splitting in the $d^0$ configuration [14]. The leading edge of the V 3$p$ peak (37.2 eV) in $Sr_4V_2O_6Fe_2As_2$ is also closer to those in $V_2O_3$ (37.8 eV) and $VO_2$ (39.5 eV) than the one in $V_2O_5$ (40.5 eV) [14]. These results suggest that the valence of V is near 3+ or 4+, which means that the V 3$d$ orbitals are partially filled and will contribute to the valence band (VB). To clarify the nature of the V 3$d$ states, we performed photon energy dependence measurements of the VB at the normal-emission direction. As seen in Fig. 1(b), there is a broad peak centered at 1 eV that is not observed in other pnictides. As noticed before, photoemission spectra of many vanadium oxides exhibit a broad peak around 1 eV, which has been attributed to the LHB due to strong correlations between V 3$d$ electrons [14-17]. The 1-eV peak height shows a different photon energy dependence compared with another prominent peak at 0.32 eV, suggesting that they have different origins. To quantitatively compare their photon energy dependence, we extracted the spectral intensity of the two peaks by using the expression $I = I_A + I_B + I_{bg}$. Here, $I_A$ and $I_B$ represent the intensity of the two asymmetric peaks at 1 and 0.32 eV, respectively, using the Doniach-Sunjic lineshape, and $I_{bg}$ is a Shirley-type background. All the spectra are well fitted within the [0.3 eV, 2 eV] range, except for those near $h\nu$ = 53 eV, where the 0.32-eV peak is strongly suppressed. The intensities of the two peaks obtained from the fitting curves are plotted as a function of photon energy in Fig. 1(c). We also plot in Fig. 1(c) the photon energy dependence of the spectral intensity at $E_F$. Both the 0.32-eV peak and the $E_F$-intensity show a remarkable Fe 3p-3d anti-resonance dip around 53 eV, as observed in $Ba_{0.6}K_{0.4}Fe_2As_2$ [18], suggesting that the structures at 0.32 eV and near $E_F$ are mostly from Fe 3d orbitals. As the photon energy reaches the V 3p-3d absorption threshold of 38 eV, the 1-eV peak shows an enhancement with a Fano-like resonance profile, suggesting that it corresponds mostly to the incoherent part of V 3d states [16]. We further compare the spectra measured at 80 and 53 eV in Fig. 1(d). At $h\nu$ = 53 eV where the Fe-3d intensity is strongly suppressed, no dispersive feature that could be attributed to the coherent V 3d band is observed in the vicinity of $E_F$. Instead, there is only a broad 1-eV peak which is likely the incoherent V 3d peak, indicating that V 3d orbitals are in a Mott-insulating state.

After clarifying the insulating nature of the V 3d states, we turn to resolve the low-energy band dispersion. Figures 2(a)-2(c) show ARPES spectra measured along ΓM, ΓX and MX in the second BZ, respectively, where the spectral intensity is found to be much enhanced as compared to the one in the first BZ. While the 1-eV peak of the incoherent V 3d orbitals show little dispersion, the lower-energy features are strongly dispersive. To better track the band dispersion, we plot in Fig. 2(e) the intensity of second derivatives of the energy distribution curves (EDCs) along the high symmetry lines Γ-X-M-Γ. To overcome the failure of simple LDA calculations in predicting the Mott insulating V bands, we adopt a simple LDA+U approach to study this material, in which we only consider the correlation effects of V 3d electrons as an effective Hubbard energy U. We found that the V 3d bands are obviously pushed away from $E_F$ as U is increased, and develop an insulating gap for U ≥ 4 eV. When U reaches 6 eV, the LHB is around 1 eV similar to what is observed experimentally. We plot the LDA+U results using experimental lattice parameters with the optimized internal coordinates of As in Fig. 2(e), and further renormalize the Fe 3d bands by a factor of 1.6 to match the overall band dispersion below 1 eV. In particular, the highly dispersive band with its bottom at M is well reproduced by the renormalized calculations. While the renormalized band calculations fit the measured bands well at high $E_B$, we find discrepancies at lower energy. The near-$E_F$ band can be well reproduced by renormalizing the Fe $3d_{xy}$ band (x and y axes along ΓM) by a factor of 3.3, indicating energy or orbital dependent band renormalization effects similar to what is observed in $Ba_{0.6}K_{0.4}Fe_2As_2$ [18], possibly due to the correlation effect of Fe $3d$ electrons.

In order to resolve the detailed structure of the bands in the vicinity of $E_F$, we have performed fine-step ARPES measurements within 0.5 eV below $E_F$. Figure 3 shows the EDC plots [(a) and (d)], the intensity plots [(b) and (e)], and the second derivative plots [(c) and (f)] of ARPES spectra in the vicinity of Γ and M, respectively. As seen in Figs. 3(a)-3(c), two hole-like bands at Γ are observed. While the inner Γ-centered band barely crosses $E_F$, the outer one forms a small hole-like FS. Figures 3(d)-3(f) display clearly one electron-like band around M. From the dispersive band, a Fermi velocity of 0.57 ± 0.05 eVÅ is obtained, which is 2.7 times smaller than the value extracted from the nonrenormalized LDA+U calculations.

Figure 4(a) shows the ARPES intensity plot at $E_F$ at $h\nu$ = 80 eV obtained by assuming a fourfold symmetry for the full BZ. We clearly observe one small hole pocket at Γ and two electron pockets at M. In addition, another large hole pocket centered at Γ is observed by using the He I$\alpha$ line, as shown in Fig. 4(b), in agreement with our LDA+U calculations. The absence of that band at $h\nu$ = 80 eV [Figs. 3(a)-3(c)] is most likely due to matrix element effects. We summarize in Fig. 4(c) the four observed FS pockets by using the $k_F$ points extracted from the momentum distribution curves (MDCs) and obtained from the four-fold symmetry operations. While the hole-like $\alpha$ and $\beta$ FSs centered at Γ have areas of 0.7 and 18% of the BZ, the electron-like $\gamma$ and $\delta$ FSs at M have areas of 5 and 13%, respectively. According to Luttinger theorem on two-dimensional FS sheets which are well expected given the superlattice-like structure in this material, the four FS sheets correspond to a valence of Fe ions close to 2+, like other parent iron pnictides. From the valence of Fe ions, it is inferred that V has a "3+" valence state, consistent with the core level measurement. As shown in Fig. 4(c), the $\beta$ FS, when shifted by the (π, 0) wave vector, overlaps well with the $\delta$ FS. Such a FS topology is very similar to those observed in other optimally doped SC pnictides. Since the superconductivity in many pnictides is thought to be related to the (π, 0) interband scattering between the hole and the electron pockets, the similar fermiology observed in this different system implies the universality of pairing interactions in the SC pnictides.

Although the fermiology in $Sr_4V_2O_6Fe_2As_2$ is qualitatively similar to those of other pnictides, we observe some marked differences. $Sr_4V_2O_6Fe_2As_2$ shows very large FSs and the nesting occurs only between the outer hole and electron pockets. Many band calculations have pointed out that the relative positions of the low-energy bands as well as the fermiology are very sensitive to the As height from the Fe plane ($h_{As}$). The $h_{As}$ in $Sr_4V_2O_6Fe_2As_2$ is increased by 5% compared with that in $SrFe_2As_2$ [7, 19]. According to the band calculations [20, 21], such an increase will reduce the hopping integral for the $3d_{xy}$ orbital via the Fe-As bond, which pushes up the $3d_{xy}$ band at Γ. Therefore, the large hole pocket in $Sr_4V_2O_6Fe_2As_2$ may originate from the Fe $3d_{xy}$ orbital due to the increase of $h_{As}$ [22]. Assuming that the outer electron-like FS originates from the Fe $3d_{xy}$ orbital [20, 21], the nesting closely related to the superconductivity may be intra-$d_{xy}$-orbital in $Sr_4V_2O_6Fe_2As_2$.

Unlike the increase of $h_{As}$ in $Sr_4V_2O_6Fe_2As_2$, in another 42622 compound, $Sr_4Sc_2O_6Fe_2As_2$, the $h_{As}$ is even slightly smaller than that in $SrFe_2As_2$ [19, 23]. The variation in $h_{As}$ is naturally attributed to the strain from the perovskite-like intercalated block, which is controlled by the radius of the transition metal ions. In contrast with SC $Sr_4V_2O_6Fe_2As_2$, $Sr_4Sc_2O_6Fe_2As_2$, in which the $Sr_3Sc_2O_6$ intercalated block is considered as a band insulator, shows a semiconducting behavior [23, 24]. Such difference may arise from significant difference in the fermiology sensitive to $h_{As}$. It is noticed that applied pressure leads to the appearance of superconductivity in $BaFe_2As_2$, which may be understood based on the change of orbital characters of the quasi-nesting FSs due to structural changes [25]. Since the $h_{As}$ in 42622 can be tuned easily by changing the transition metal elements, the 42622 system provides a good opportunity to study by ARPES the relationship among superconductivity, electronic structure and crystal structure.

Remarkably, the spectral line shape of $Sr_4V_2O_6Fe_2As_2$ is quite different from those of many other pnictides. As seen in Fig. 1(d), the intensity near $E_F$ is suppressed strongly compared with the VB, whereas the $E_F$-intensity is comparable with the VB in $BaFe_2As_2$. Interestingly, the suppression of spectral intensity only occurs within ~ 0.1 eV below $E_F$, while the band renormalization near $E_F$ and the Fermi velocities are not significantly different from those of $Ba_{0.6}K_{0.4}Fe_2As_2$ [18]. This is reminiscent of the photoemission results in $CaVO_3$ and $SrVO_3$, where the near $E_F$-intensity of $CaVO_3$ is strongly suppressed while its band width is comparable to that of $SrVO_3$ [17]. This has been attributed to a strong momentum dependence of the self-energy as long-range interactions become important in the vicinity of a Mott transition [17]. Therefore, our ARPES results suggest that the itinerant Fe 3$d$ states in $Sr_4V_2O_6Fe_2As_2$ may be pushed toward a Mott transition with the insertion of the Mott insulating $Sr_3V_2O_6$ layers, while the microscopic origin of the possible long-range interactions remains to be clarified in future experimental and theoretical studies.

This work was supported by grants from CAS, NSFC, MOST of China, JSPS, TRIP-JST, CREST-JST, MEXT of Japan, and NSF, DOE of US. This work is based upon research conducted at the SRC supported by NSF DMR-0537588 and the PF supported by PF-PAC 2009S2-005.

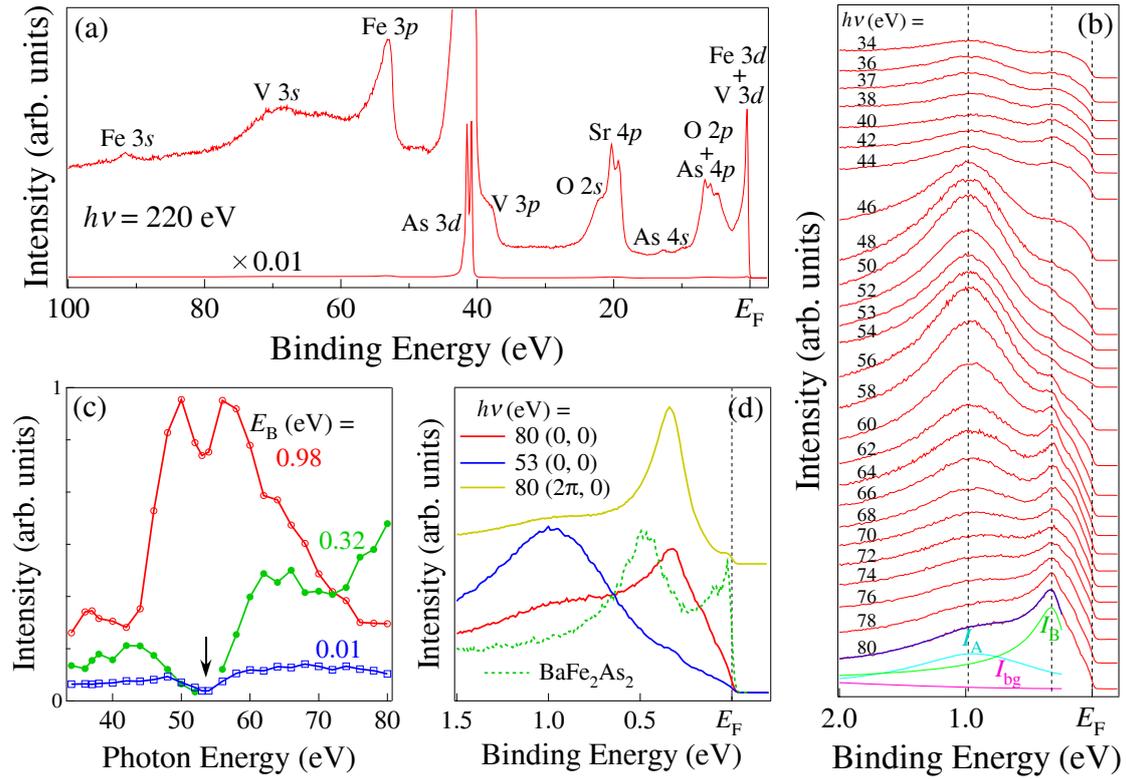

Fig. 1: (a) Core level photoemission spectrum at $h\nu = 220$ eV. Associated atomic orbitals are indicated. To show completely the strong peaks from As $3d$ orbital, the spectrum divided by 100 is also plotted. (b) VB at BZ center measured at different photon energies (34 – 80 eV). All the spectra are normalized by the photon flux. The fitting curves for the spectrum at $h\nu = 80$ eV are also shown. (c) Photon energy dependence of the intensities of the two peaks obtained from the fitting curves and the spectral intensity at $E_B = 0.01$ eV. (d) Direct comparison of the VB measured at $h\nu = 53$ and 80 eV, along with the VB of BaFe$_2$As$_2$ ($h\nu = 80$ eV).

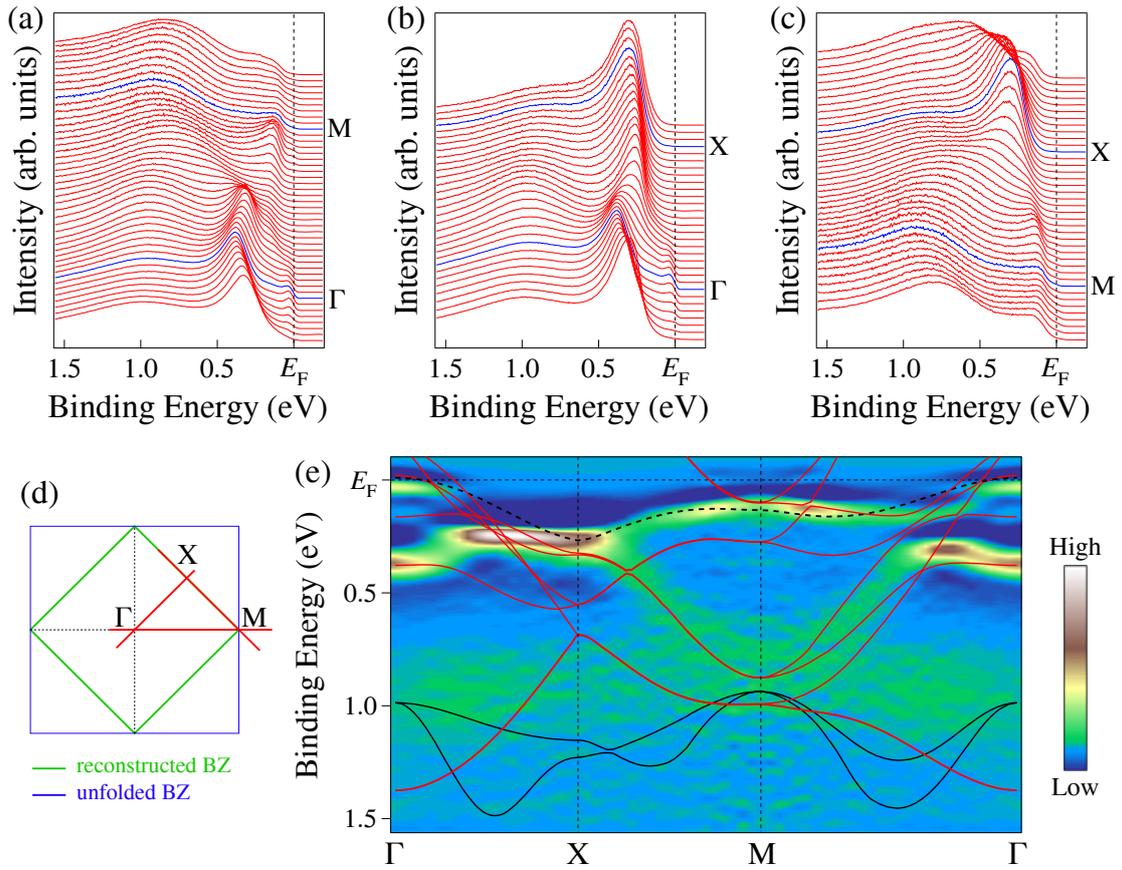

Fig. 2: Dispersive EDCs along ΓM (a), ΓX (b) and MX (c) measured at $h\nu$ = 80 eV. (d) Schematic BZ indicating measurement locations for panels (a)-(c). (e) Second derivative plot along ΓXMΓ. LDA+U bands are also potted for comparison. The Fe 3$d$ bands (red lines) are renormalized by a factor of 1.6, whereas the V 3$d$ bands (black lines) are not. The Fe 3$d_{xy}$ band (black dashed line) is renormalized by a factor of 3.3 to reproduce the experimental band near $E_F$.

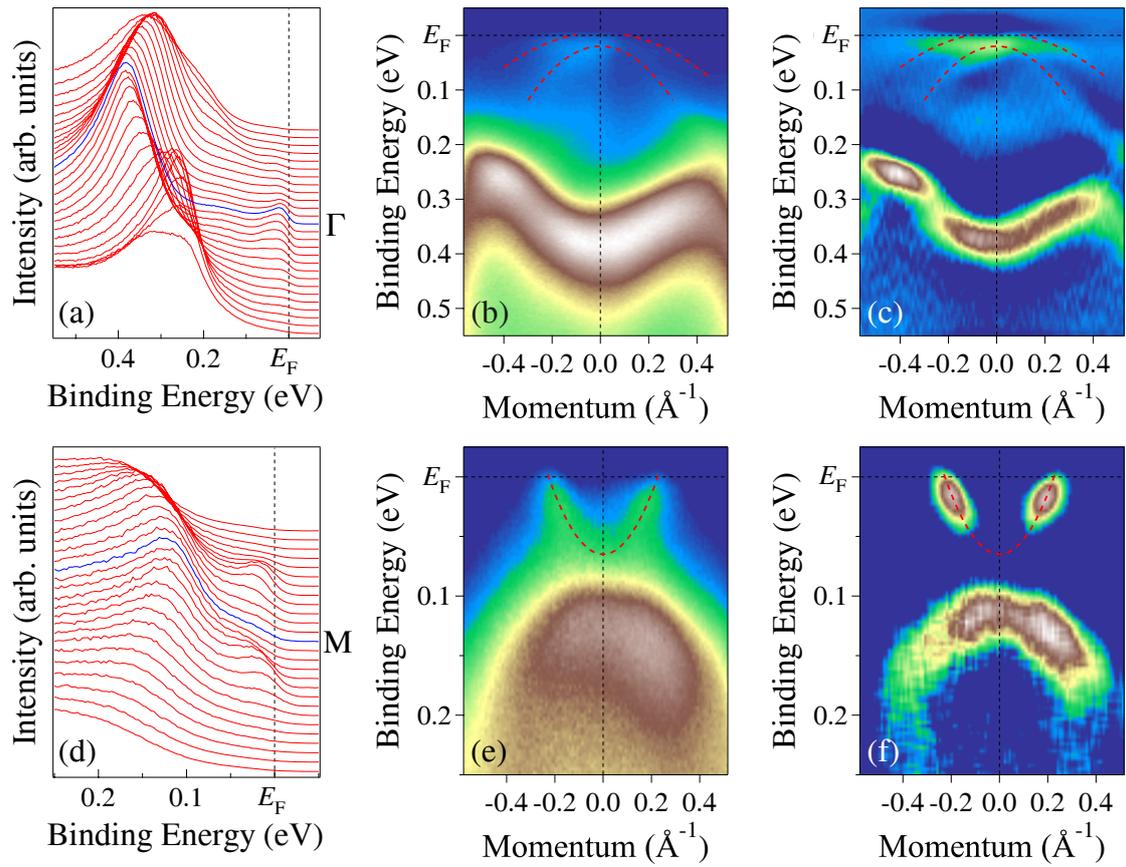

Fig. 3: EDCs (a), intensity plot (b) and second derivative plot (c) in the vicinity of $E_F$ at Γ measured at $h\nu$ = 80 eV. (d)-(f) same as in panels (a)-(c), but taken at M. The second derivatives of EDCs are used in panel (c). We summed the second derivatives of EDCs and MDCs in panel (f) to better track the fast dispersive band near $E_F$ at M. Red dashed lines are guides for eyes.

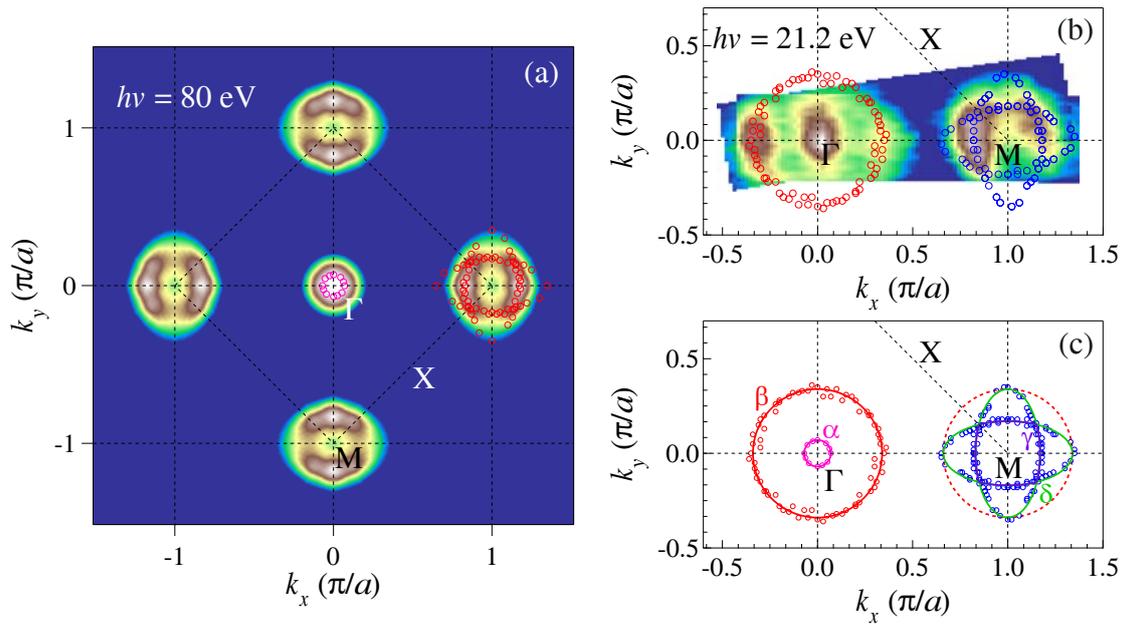

Fig. 4: ARPES intensity plot at $E_F$ as a function of the two-dimensional wave vector measured at $h\nu = 80$ (a) and 21.2 eV (He I$\alpha$ line) (b). The intensity at $E_F$ is obtained by integrating the spectra within $\pm 20$ meV with respect to $E_F$. Open circles show experimentally determined $k_F$ points symmetrized by assuming a fourfold symmetry with respect to $\Gamma$ and M. (c) Extracted FSs (solid lines) from the extracted $k_F$ points. The dashed FS is the $\beta$ FS shifted by the $(\pi, 0)$ wave vector.